\documentclass{aa}
\usepackage[varg]{txfonts}
\usepackage{graphicx,xcolor}
\usepackage[colorlinks=true,citecolor=blue,linkcolor=blue]{hyperref}
\usepackage[export]{adjustbox}

\begin{document}

\title{Readdressing the UV solar variability with SATIRE-S: non-LTE effects}
\titlerunning{SATIRE-S with non-LTE spectra}

\author{R.~V.~Tagirov\inst{1,2},
        A.~I.~Shapiro\inst{2},
        N.~A.~Krivova\inst{2},
        Y.~C.~Unruh\inst{1},
        K.~L.~Yeo\inst{2}
        \and
        S.~K.~Solanki\inst{2, 3}}
\authorrunning{R.~V.~Tagirov et al.}

\institute{Blackett Laboratory,
           Astrophysics Group,
   		   Imperial College,
           London SW7 2AZ, UK\\
           \email{rtagirov@imperial.ac.uk}
           \and
           Max-Planck Institute for Solar System Research,
           37077 G{\"o}ttingen,
           Germany
           \and
           School of Space Research, 
           Kyung Hee University, 
           446-701 Yongin, 
           Gyeonggi-Do, 
           Korea}

\date{}
\abstract
{Solar spectral irradiance (SSI) variability is one of the key inputs to models of the Earth's climate.
Understanding solar irradiance fluctuations also helps to place the Sun among other stars
in terms of their brightness variability patterns and to set detectability limits for terrestrial exo-planets.}
{One of the most successful and widely used models of solar irradiance variability is SATIRE-S. It uses spectra of the magnetic features and surrounding quiet Sun computed with the ATLAS9 spectral synthesis code under the assumption of Local Thermodynamic Equilibrium (LTE).
SATIRE-S has been at the forefront of solar variability modelling, but due to the limitations of the  LTE approximation its output SSI has to be empirically corrected below 300 nm, which reduces the
physical consistency of its results.
This shortcoming is addressed in the present paper.}
{We replace the ATLAS9 spectra of all atmospheric components in SATIRE-S with 
the spectra calculated using the non-LTE Spectral Synthesis Code (NESSY). We also use  \cite{fontenla1999} temperature and density stratification models of the solar atmosphere to compute the spectrum of the quiet Sun and faculae.}
{We compute non-LTE contrasts of spots and faculae and combine them with 
the SDO/HMI filling factors of the active regions 
to calculate the total and spectral solar irradiance variability during solar cycle 24.
}
{The non-LTE contrasts result in total and 
spectral solar irradiance in good agreement with the empirically corrected output of the LTE version.
This suggests that empirical correction introduced into SATIRE-S output is well judged and that the corrected total and spectral solar irradiance obtained from the SATIRE-S model in LTE is fully consistent with the results of non-LTE computations.}


\maketitle

\section{Introduction}
The variability of the solar spectral irradiance is one of the drivers of the chemistry
and dynamics in the Earth’s middle atmosphere \citep{brasol2005, haigh2007} and
may affect the terrestrial climate 
\citep{haigh1994, souhoo2006, austin2008, haigh2010, solanki2013, mitchell2015}.
To understand the nature and the degree of this influence, and thus disentangle it from the anthropogenic causes of climate change, chemistry-climate models have been developed in which solar irradiance is a key input component.

Studies of solar irradiance variability also help to understand the variability of other stars, while
a comparison of the solar variability pattern to that of other lower main sequence stars helps to
understand the place of the Sun among them \citep[e.g.,][]{basri2013, shapiro2013b, shapiro2014}.
Furthermore, stellar variability is a limiting factor for the detection of terrestrial exoplanets 
\citep{saadon1997, aigrain2004, desort2007, pont2008, lagrange2010, gilliland2011},
which calls for realistic models of stellar brightness variations 
\citep{boisse2012, korhonen2015}.
Note that the extension of the exoplanetary search mission Kepler \citep{batalha2014} 
was granted due to higher than expected stellar variability \citep{gilliland2015}.

Gaining an understanding of the solar-terrestrial and solar-stellar connections requires sufficiently long, uninterrupted and reliable records of solar spectral irradiance.
The available measurements fulfill these requirements only in part.
The Total Solar Irradiance (TSI), which is the spectrally integrated radiant flux 
received per unit surface area at 1 AU from the Sun, 
has been monitored almost continuously by a series of missions since 1978 \citep[see, e.g.,][]{kopp2016}.
The measurements of Spectral Solar Irradiance (SSI), i.e. radiant flux per unit surface area and unit wavelength at 1 AU from the Sun, are significantly less homogeneous.
The records have numerous gaps both in time and wavelength, and they exhibit significant instrumental trends \citep[see][]{yeo2015}.
Thus models of the solar irradiance variability have been developed to allow 
reconstructions with a more complete coverage both in time and wavelengths \citep[see, e.g.,][for a review]{solanki2013}.

The first generation of solar irradiance models relied on empirical 
correlations between the measured irradiance and proxies of solar magnetic activity.
The more advanced semi-empirical models used 1D representations of temperature and density 
stratifications of active and quiet components of the Sun (solar model atmospheres) and radiative transfer codes to calculate their spectra.
This second generation of models, having at its core a more robust physical approach,
provided insights into understanding of the physics behind the empirical correlations. In particular,
\cite{krivova2003}, \cite{ball2012} and \cite{yeo2014} have shown that over 96\% of TSI variations during cycles 23 and 24 on timescale of a day and longer can be explained by the evolution of the magnetic field at the solar surface alone.

Among the semi-empirical models,
the Spectral And Total Irradiance Reconstruction model \citep[SATIRE-S,][]{fligge2000, krivova2011b} 
is one of the most successful.
The index ``S'' stands for the version of the model based on satellite data, i.e. full-disc continuum images and magnetograms.
SATIRE-S employs the Local Thermodynamic Equilibrium (LTE) radiative transfer 
code ATLAS9 \citep{kurucz1992a, caskur1994} for calculation of spectra emergent 
from the active (i.e. strongly magnetized) and quiet regions of the Sun.
Because of this LTE limitation the SSI calculated with SATIRE-S is subject to empirical correction below 300~nm \citep{krivova2006, yeo2014}. 
Thus, in the range between 180~nm and 300~nm, the absolute level of the irradiance is shifted to match that of the WHI spectrum \citep{woods2009}, while below 180 nm also the magnitude of the rotational and solar-cycle variability is scaled to agree with SORCE/SOLSTICE measurements \citep{mcclintock2005}.
Although the results of this scaling provide rather accurate TSI  and satisfactory SSI reconstructions, such an ad hoc step is unsatisfactory. In addition, there are strong spectral lines in this wavelength range, whose variability is poorly reproduced in LTE, irrespectively of the scaling.

Unfortunately, this weakness of SATIRE-S is particularly pronounced in 
the most controversial spectral range from 200~nm to 400~nm. 
The irradiance variability in this range is important for climate studies 
\citep{ermolli2013, ball2014b, ball2014a, ball2016, maycock2015,maycock2016}. 
It is thus vexing that the magnitude of the variability in this range differs by a factor of several among the different models and the measurements \citep{krivova2006, harder2009, unruh2012, yeo2015, yeo2017a}.
Empirical models tend to return variability amplitudes 
that are almost a factor of two lower than semi-empirical models suggest 
\citep{thuillier2014b, yeo2014, yeo2015, yeo2017a, egorova2018}, while the insufficient stability of most instruments \citep{yeo2015} compared to the solar cycle variability above 250~nm - 400~nm does not allow 
deciding between these different amplitudes.

In this paper we use the non-LTE Spectral Synthesis (NESSY) code \citep{tagirov2017} to compute the emergent spectra of the various atmospheric components used in SATIRE-S and to model SSI variability over cycle 24 without the need for any empirical corrections.
We emphasize that the aim of this paper is not to update the SATIRE-S model, 
but rather to compare LTE and non-LTE active region contrasts and to verify the model
in the UV part of the solar spectrum. 

In Sect.~\ref{desc} we give a description of SATIRE-S and the verification procedure, 
in Sect.~\ref{res} we discuss the results and in Sect.~\ref{conc} we summarize our 
results and provide conclusions and outlook.

\section{Model Description}\label{desc}
The SATIRE-S model returns daily values of solar irradiance 
in the wavelength range from 115 nm to 160000 nm.
The model operates under the assumption that the solar irradiance variability on timescales longer than a day is driven by the magnetic active regions on the surface of the Sun and
has two main components: the fractional coverages of the solar disc by these regions 
and their synthesized intensity spectra.
The active regions included in the model are spots (consisting of umbra and penumbra components), which reduce solar irradiance as they pass across the solar disc, 
and faculae, which have the opposite effect (at wavelengths up to about 1500~nm).
SSI, as a function of time $t$ and wavelength $\lambda$, can be written as
\begin{gather}
\nonumber
S(t, \lambda) = S_q(\lambda) + \sum_{ji}\alpha_{ji}(t)C_j(\lambda, \mu_i)\Delta\Omega_i,\\
\nonumber
S_{q}(\lambda) = \sum_iI_{q}(\lambda, \mu_i)\Delta\Omega_i,\\
\label{irr_eq}
C_j(\lambda, \mu_i) = I_j(\lambda, \mu_i) - I_q(\lambda, \mu_i),
\end{gather}
where $j \in \{f, p, u\}$ and $I_q$, $I_f$, $I_p$, $I_u$ are the intensities emerging
from quiet Sun, faculae, penumbra and umbra respectively.
The intensities are functions of wavelength and of the cosine of heliospheric angle $\mu_i$. 
The solar disc has been subdivided into concentric rings (indexed with $i$) covering an equally large $\mu$ range.
The integration over the solar disc is represented by the summation over the concentric rings.
Furthermore, $\alpha_{ji}$ are the time-dependent fractional coverages (or filling factors) of the $i^{\mathrm{th}}$ ring by the faculae, penumbrae and umbrae,
$C_j(\lambda, \mu_i)$ is the brightness contrast of magnetic feature $j$,
$\Delta\Omega_i$ is the solid angle extended by the $i^\mathrm{th}$ ring as seen from the Earth and
$S_q(\lambda)$ is the full disc quiet Sun radiative flux at wavelength $\lambda$.

In SATIRE-S, the surface coverages by faculae and sunspots are derived from the continuum intensity 
images and magnetograms taken by NSO/KP, SOHO/MDI and SDO/HMI \citep[see][for more details]{yeo2014}.
Here we only consider the period from 2010 to 2017 
covered by the superior space-borne SDO/HMI observations.
We stress once more that our aim is to validate SATIRE's variability in the UV range, rather than a full update of the model.

The filling factors are given by the ratio of the area (in pixels) covered 
by a particular type of feature (faculae,  sunspot umbrae or penumbrae) 
to the total area of the $i^\mathrm{th}$ ring.
Both spots and faculae are seen in magnetograms, but only spots are seen in the intensity images as dark regions.
Therefore, by combining these two types of images one can deduce whether a given pixel has facular signal.
Faculae are composed of a multitude of magnetic elements, whose sizes are such that they do not 
always fill the area of an entire pixel.
If a pixel is magnetic and does not belong to a spot, then a fraction of this pixel
is attributed to faculae.
This fraction is linearly proportional to the magnetic signal coming from the pixel up to a saturation value $B_\mathrm{sat}$ \citep{fligge2000}, above which the pixel is assumed to be completely covered by faculae.
$B_\mathrm{sat}$ is a free parameter in the model and its optimal value 
is determined by varying it until best agreement is reached between modelled and measured TSI variability \citep{yeo2014}.
Note, that a new-generation version of the SATIRE-S model, SATIRE-3D, has recently been developed, which no longer needs this free parameter \citep{yeo2017b}. However, this version currently deals with TSI only, 
so that here we consider the older version of SATIRE-S that still requires $B_\mathrm{sat}$.

The most recent SATIRE-S spectral solar irradiance dataset \citep{yeo2014} 
employs intensity spectra of solar surface features calculated by \cite{unruh1999}, henceforth referred to as U99, 
using the radiative transfer code ATLAS9, which operates under the assumption of LTE.
The U99 spectra have been produced with resolution $\Delta\lambda = \{1, 2, 5\}$ nm 
in the $\{100 - 290, 290 - 1000, 1000 - 1200\}$ nm intervals, respectively. 
The quiet Sun, umbra and penumbra features  {have been} described by
stellar atmosphere models from the Kurucz grid
\citep{kurucz1993}, in which atmospheric temperature and density stratifications are specified by effective temperature $T_\mathrm{eff}$ and surface gravity $\log g$. 
The models with $(5777\ \mbox{K},\ 4.44)$, $(5400\ \mbox{K},\ 4.0)$ and $(4500\ \mbox{K},\ 4.0)$ are used to represent temperature and density stratifications in the quiet Sun, penumbra and umbra, respectively; a version of the FAL93-P model of \cite{fontenla1993} modified by U99 is used to represent the facular stratification.

\begin{figure}
\centering
\includegraphics[width=9.15cm]{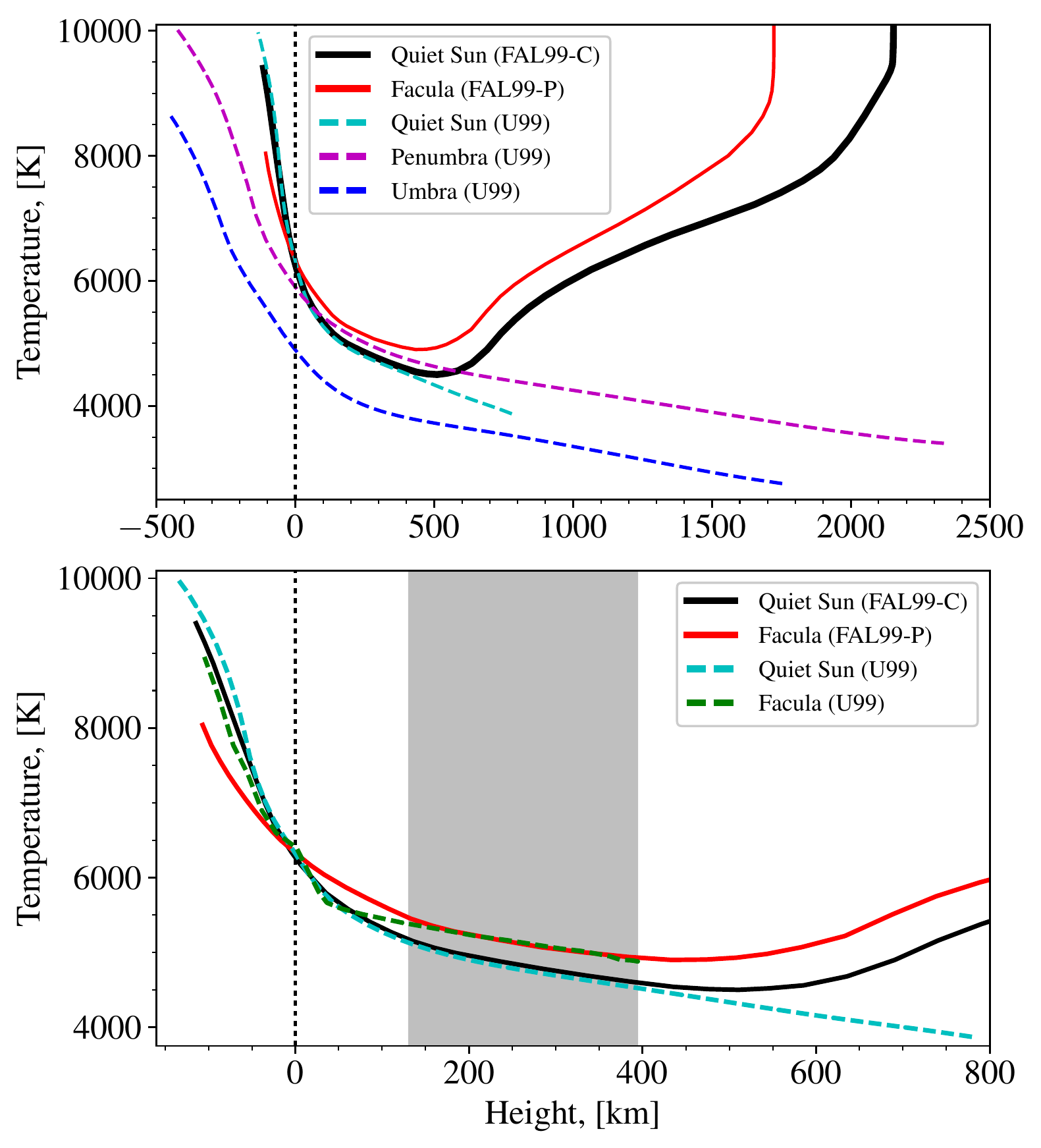}
\caption[]{ {Top panel:}    Temperature stratification in the atmospheric models used for the non-LTE version of SATIRE-S.
                    The FAL99-C and FAL99-P models \citep{fontenla1999} are shown as solid lines and were used to calculate the facular contrast in Eq.~\eqref{irr_eq}. 
                                 {The dashed lines show Kurucz solar and stellar models (see U99) used to derive spot contrasts.}
          {Bottom panel:} Comparison of the temperature stratifications used 
                                in LTE and non-LTE versions of SATIRE-S
                                for the calculation of the facular contrast.
                                The dashed green curve was derived by modifying the FAL93-P model of \cite{fontenla1993}.
                                The height grids of all models in both panels were offset 
                                so that the zero height point in each model
                                corresponds to Rosseland optical depth  {(calculated in LTE)} equal to 2/3.
                                 {The gray area marks the region of equal temperature difference between quiet Sun and facula in the U99 and FAL99 temperature stratifications 
                                (see the Appendix for discussion).}
                                }
\label{fig:atm_mod}
\end{figure}

The models in the U99 set do not include any chromosphere 
which is necessary to avoid artefacts such as strong emission lines when the spectrum is calculated in LTE. 
However, it is insufficient for the purposes of this paper, 
since the SSI below 200~nm cannot be modeled correctly without chromospheric layers of the quiet Sun and faculae.
Therefore, we use the atmospheric models FAL99-C and FAL99-P 
\citep{fontenla1999} to compute the $S_q(\lambda)$ term and the facular contrast $C_f$ in  Eq.~\eqref{irr_eq}.
At the same time, we keep the 
 {Kurucz} atmospheric models to calculate 
the penumbral and umbral contrasts ($C_p$ and $C_u$ in Eq.~\ref{irr_eq}, respectively).
 {Kurucz models} give a more realistic effective 
temperature difference between the quiet Sun and umbra 
$\Delta T^{qu}_\mathrm{eff} = 1277\ \mathrm{K}$ 
than the FAL99 set with $\Delta T^{qu}_\mathrm{eff} = 1887\ \mathrm{K}$,
which is very close to the upper edge of the observed $\Delta T^{qu}_\mathrm{eff}$ 
range \citep[see, e.g.,][p. 161]{solanki2003} and hence is applicable only to a small subset of sunspots.
Consequently, the spot contrast  {calculated with the Kurucz models} leads to a more accurate solar irradiance variability in the visible and infrared where spots have significant contribution to  {it}.
For sunspots, the continuum plays a much more important role than the lines, and, moreover, below 200~nm the spot contrast contributes much less to TSI and SSI than the facular contrast. 
Therefore, the absence of a chromosphere in U99 umbral and penumbral models has minimum effect. 
In any case, the chromospheres of commonly used umbral and penumbral models are not consistent with the constraints recently imposed by observations made with the Atacama Large Millimeter Array \citep{loukitcheva2017}.
The resulting set of atmospheric models used in our calculations is shown in the top panel of Fig.~\ref{fig:atm_mod}. The comparison of the quiet Sun and facular models from FAL99 and U99 sets is shown in the bottom panel.

The latest version of the NESSY code \citep{tagirov2017} has been used to compute the non-LTE intensities in Eq.~\eqref{irr_eq}.
The code solves the 1D spherically symmetric non-LTE radiative transfer problem 
for a given temperature and density stratification.
The radiative transfer equation and the system of  {Statistical Balance Equations (SBEs)} are solved simultaneously for all chemical elements from hydrogen to zinc.
The atomic model for each element consists of the ground and first ionized states with varying number of energy levels in the ground state and one energy level in the ionized state.
Overall, the atomic model of the code has 114 energy levels  {(of which 30 are first ionized states)}, 217 bound-bound transitions and 84 bound-free transitions across the elements included in the model.
Once the non-LTE problem is solved, the spectral synthesis block of the code takes 
the non-LTE energy level populations and employs a linelist compiled from the Kurucz linelist (pers. comm.) and the Vienna Atomic Line Database 
\citep{kupka1999, kupka2000} to compute the high resolution  {(2000 points per nm)} spectrum. We have adopted a value of 1.5 km/s for the microturbulent velocity during spectral synthesis calculations, which is the same value as was used by \cite{unruh1999} for facular and quiet Sun spectra.  We note that the resulting spectra and, consequently, contrasts of magnetic features depend on the adopted value.

\begin{figure*}[t]
\centering
\includegraphics[width=18.5cm]{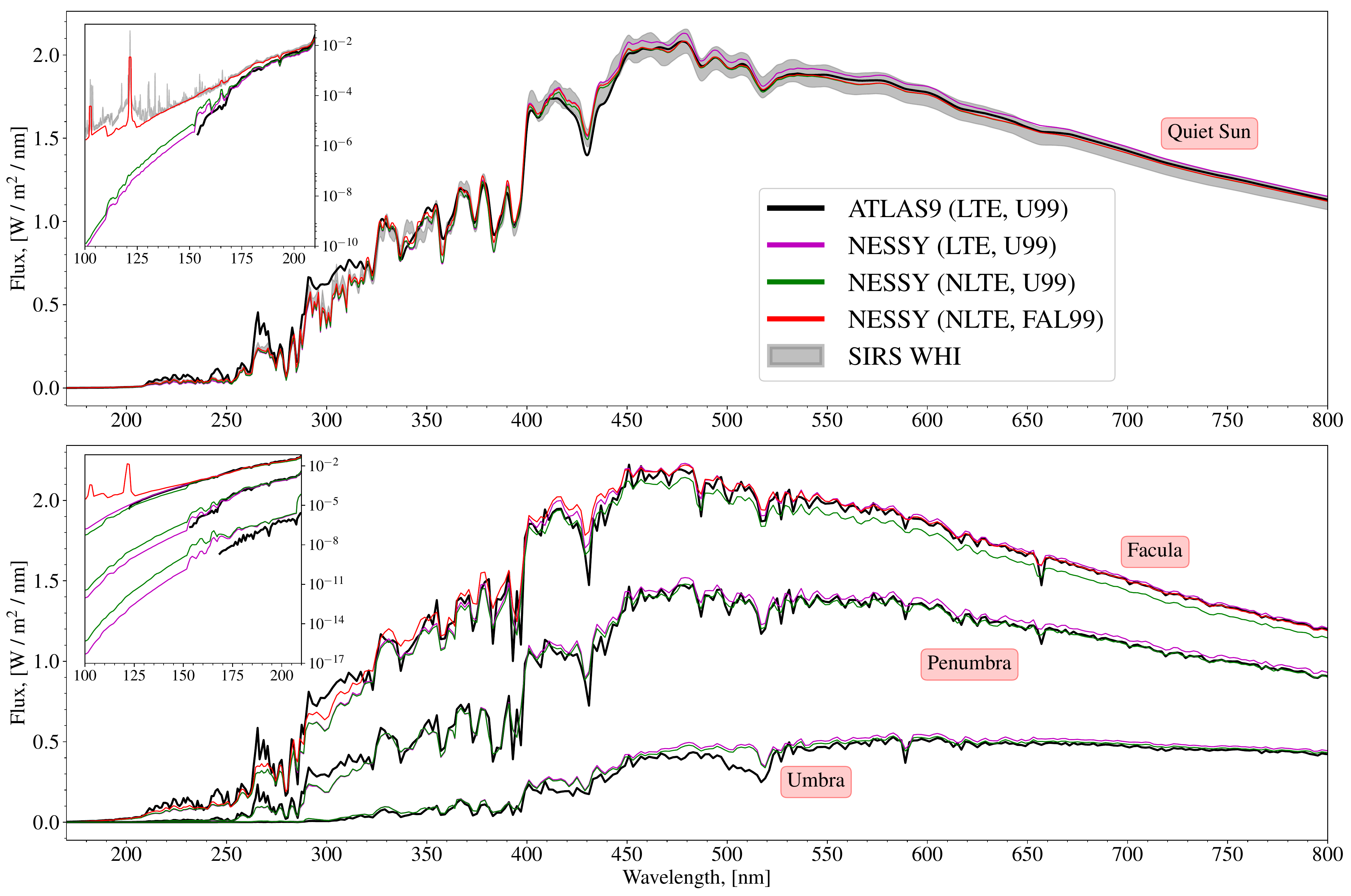}
\caption[]{ {Top panel:} 
 {Comparison of NESSY and ATLAS9 quiet Sun spectra to SIRS WHI observations \cite{woods2009}.
NESSY spectra are calculated for three cases. 
Each case highlights the change in the spectrum at every single step in the progression from ATLAS9 LTE calculation with the set of U99 models to NESSY non-LTE calculation with the FAL99 models. The gray area designates the 1$\sigma$ uncertainty range of the observations. 
NESSY and ATLAS9 spectra were convolved with the SIRS WHI spectral resolution.}
 {Bottom panel:} 
 {Comparison of facular, umbral and penumbral spectra for the same set-ups.
The colouring of the curves corresponds to that in the top panel. The three groups of curves on the main panel and the inset correspond, from bottom to top, to umbra, penumbra and facula. NESSY spectra are binned to the ATLAS9 spectral grid.}
}
\label{fig:spec_comp}
\end{figure*}

 {The solar spectrum contains tens of millions of atomic and molecular lines \citep[see, e.g.,][]{SHS_book}. 
These lines dominate the SSI variability in the UV, violet, 
blue, and green spectral domains  \citep{shapiro2015}. 
Considering all of these lines in non-LTE is presently not feasible.}
 {Therefore, a pseudo non-LTE approach is employed in NESSY. 
In this approach, the code first solves the SBEs for all levels of the atomic model described above.
Then, the resulting non-LTE populations of the ground and first ionized states of each element, as well
as of all hydrogen levels, are passed to the spectral synthesis block of the code.
The level populations for all other transitions included in the spectral synthesis 
are calculated in pseudo non-LTE, i.e., using the Saha-Boltzmann distribution 
with respect to the non-LTE populations of the ground and first ionized states.
Such treatment allows to 
take into account the over-ionization effects, 
crucial for proper representation of 
the spectral profile, in particular in the UV 
\citep[see, e.g.,][]{shohau2009, shapiro2010, rutten2019}.
}

 {We note that molecular lines are currently treated in LTE, which, however, is a reasonable assumption since non-LTE effects in the main molecular bands of the solar spectrum are relatively small \citep[see, e.g.][]{Kleint_C2,shapiro_CN}. A more detailed description of the molecular line treatment in NESSY is given in \cite{shapiro2010}}.

 {The over-ionization effects and, consequently, 
the overall intensity distribution of the emergent 
spectrum of a star (especially in the UV) are 
influenced by the numerous bound-bound absorptions.
This effect is known as line-blanketing \cite[][p. 167]{mihalas1978}.
One of the reasons for this influence is the change of the ionization equillibrium 
state in the stellar photospheric layers
due to ionizing continuum radiation being blocked by the spectral lines.
In the non-LTE case the ionization equilibrium is governed by the SBEs. 
In NESSY the line-blanketing effect on the SBEs
is included via the LTE Opacity Distribution Function (ODF) procedure.
All in all, calculations are performed in the following four steps:
\begin{itemize}
\item Calculation of LTE populations of the levels specified in the atomic model described above;
\item ODF synthesis in the 100 nm - 1000 nm 
spectral range using the LTE populations;
\item Solution of the SBEs, i.e., calculation of non-LTE populations 
of the levels in the atomic model taking into account the ODF;
\item Spectral synthesis using the non-LTE populations.
\end{itemize}
For the spectral region considered in this paper (100 nm - 1100 nm) this 
ODF procedure gives the same results as the procedure described by \cite{haberreiter2008} and \cite{tagirov2017}. 
The described atomic model, the pseudo non-LTE approach and the ODF procedure
for taking the effects of line-blanketing on SBEs 
have been shown to be adequate for representation of the solar spectrum \citep{shapiro2010, tagirov2017}.}

\section{Results}\label{res}
\begin{figure}
\centering
\includegraphics[width=9.15cm]{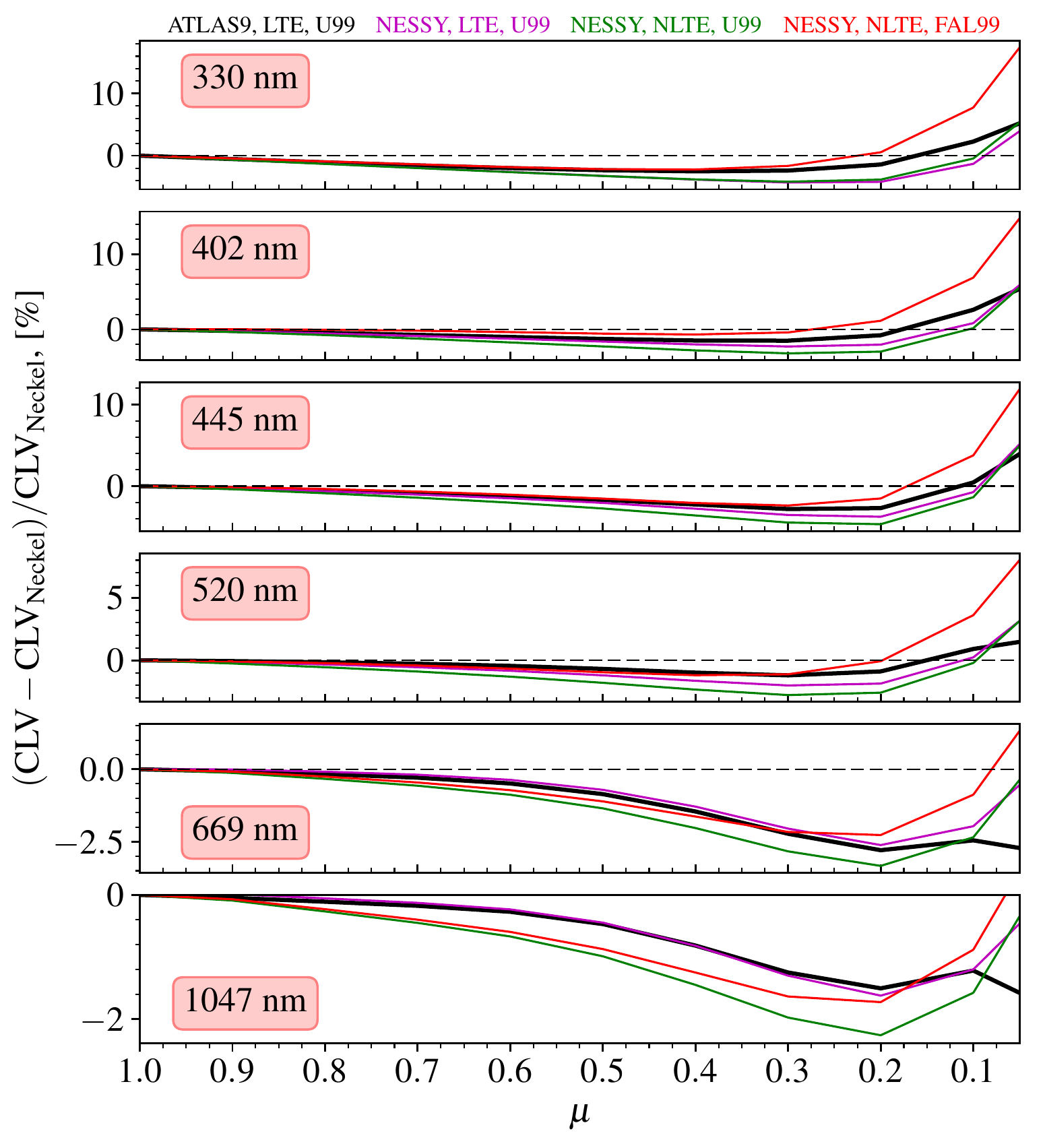}
\caption[]{Relative deviations of CLVs calculated with ATLAS9 and NESSY 
         (for the same cases as in Fig.~\ref{fig:spec_comp})
         from the CLV fits by \cite{neclab1994}.
         The wavelength selection is the same as in Fig.~2 of U99. 
         NESSY CLVs are binned to the ATLAS9 spectral grid.
         The $x$-axis is the cosine of heliospheric angle. 
         Note the different scales for the $y$-axes.
		 }
\label{fig:clv}
\end{figure}
\begin{figure}
\centering
\includegraphics[width=9.15cm]{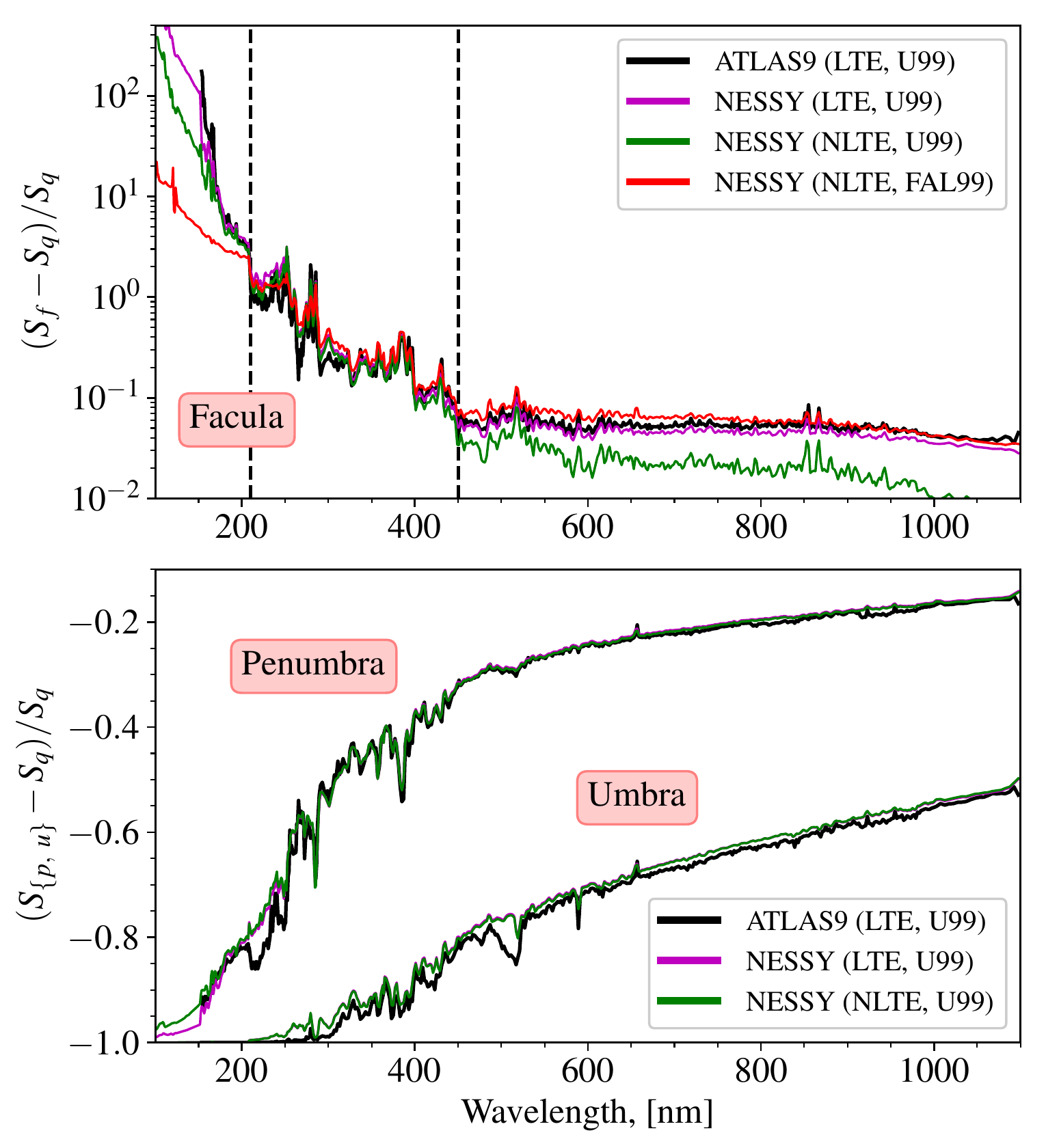}
\caption{ {Relative full-disc facular, umbral and 
         penumbral contrasts (Eq.~\ref{eq:rfdc})
         as calculated with ATLAS9 in LTE for U99 models and with NESSY in LTE and non-LTE for U99 and FAL99 models (faculae only).
         Each curve is derived from the corresponding curves in Fig.~\ref{fig:spec_comp}
         and shows the impact on the relative contrasts produced by the change of code,
         radiative transfer treatment and the atmospheric model set.
         Shortward of about 150 nm the black curves are truncated because of the limitations of the ATLAS9 code at these wavelengths.
         NESSY contrasts are binned to the ATLAS9 spectral grid.
         The dashed lines designate the spectral domain of similar facular contrasts (see main text).}
		 }
\label{fig:contr}
\end{figure}
 {Figure~\ref{fig:spec_comp} compares the quiet Sun and active component spectra calculated with NESSY to the spectra calculated with ATLAS9 and to observations.
In the SATIRE-S setup used here, the radiative transfer code, the spectral synthesis treatment, and the atmospheres used to derive facular contrasts have all changed. We thus use a sequence of colours (purple, green and red, respectively) to show the effects of these changes step by step. The black, purple and green lines in Figs.~\ref{fig:spec_comp} to \ref{fig:contr} (and in Fig.~\ref{fig:fheight}) show calculations based on the set of model atmospheres described in U99. Black and purple are for LTE calculations with ATLAS and NESSY and are denoted ATLAS-LTE-U99 and NESSY-LTE-U99 in the text. Spectra and contrasts shown in green are for NESSY calculations in (pseudo) non-LTE and are labelled NESSY-NLTE-U99. The red lines, finally, show NESSY (pseudo) non-LTE calculations for models C (quiet Sun) and P (faculae) from the FAL99 atmosphere set; these are denoted as NESSY-NLTE-FAL99.}

 {The top panel in Fig.~\ref{fig:spec_comp} compares the synthesized quiet Sun spectra to the SIRS WHI reference quiet Sun spectrum \citep{woods2009}.
The gray area is the 1$\sigma$ uncertainty range of the observations.
The calculated spectra were convolved with the SIRS WHI spectral window
\citep[the procedure is the same as in][]{tagirov2017}.}

 {The inset plot in the top part of Fig.~\ref{fig:spec_comp} shows that all spectra calculated with  {the Kurucz} quiet Sun model fall below the observed spectrum shortward of about 180~nm.
This is explained by the absence of chromospheric layers in that model.
The black curve is cut short around 150~nm for the same reason. 
Technical details of ATLAS9 radiative transfer 
numerical scheme result in extremely low intensity values for this spectral domain.
}

 {The lack of spectral lines in the NESSY-NLTE-FAL99 curves between 125~nm and 180~nm is because all lines in this spectral region, except for the hydrogen Lyman series, are treated in pseudo non-LTE 
in NESSY (see Sect.~\ref{desc}).
These spectral lines form high above the temperature minimum and when treated in pseudo 
non-LTE the amount of intensity emitted in each of them is erroneously high. 
To avoid it, in NESSY, the value of the source function for these lines is set to the value of the Planck function at the temperature minimum.
This, in turn, leads to insufficient emission when compared to observations.
A full non-LTE treatment of these lines is necessary in order for them to appear in the non-LTE spectrum.}

 {Another feature of the red curve is the underestimated Ly$\alpha$ flux.
This can be remedied by adjusting the Doppler broadening velocity parameter in the non-LTE block of NESSY \cite[see Sect.~5.2 in][]{tagirov2017}.
However, since the focus of this work was on the verification of the overall SATIRE-S SSI variability profile, such Doppler velocity tuning has not been performed.
}

 {Longward of 450~nm the NESSY-NLTE-U99 spectrum (green line) falls below the NESSY-LTE-U99 spectrum (purple) because of the non-LTE increase of H$^-$ concentration in the photospheric layers. H$^-$ is dissociated mainly by the near-IR photons \citep[][p. 102]{mihalas1978}. Increased H$^-$ concentration, caused by the visible and near-IR photons escaping the solar atmosphere, leads to a higher H$^-$ opacity.
Thus, the continuum longward of 450~nm forms at lower photospheric temperatures.
The non-LTE H$^-$ concentration effect was originally reported by \cite{vernazza1981} and has been demonstrated for the FAL99 set of models by \cite{shapiro2010}.}

 The bottom panel of Fig.~\ref{fig:spec_comp} shows the comparison of ATLAS9 and NESSY spectra of facula, umbra and penumbra.
NESSY spectra were binned to the ATLAS9 spectral grid.
The three groups of curves in the main panel and inset correspond to umbra (bottom), penumbra (middle) and facula (top), respectively.
For wavelengths above 450~nm, the  aforementioned H$^-$ effect is seen in all three cases. The noticeably lower flux in ATLAS9 umbra calculations between 490~nm and 530~nm is attributed to the opacity from MgH bands, which is not taken into account in NESSY calculations but is significantly overestimated in the old linelists used by \cite{unruh1999} for their ATLAS9 calculations \citep[see, e.g.][]{weck2003}.

\begin{figure*}
\centering
\includegraphics[width=18.5cm]{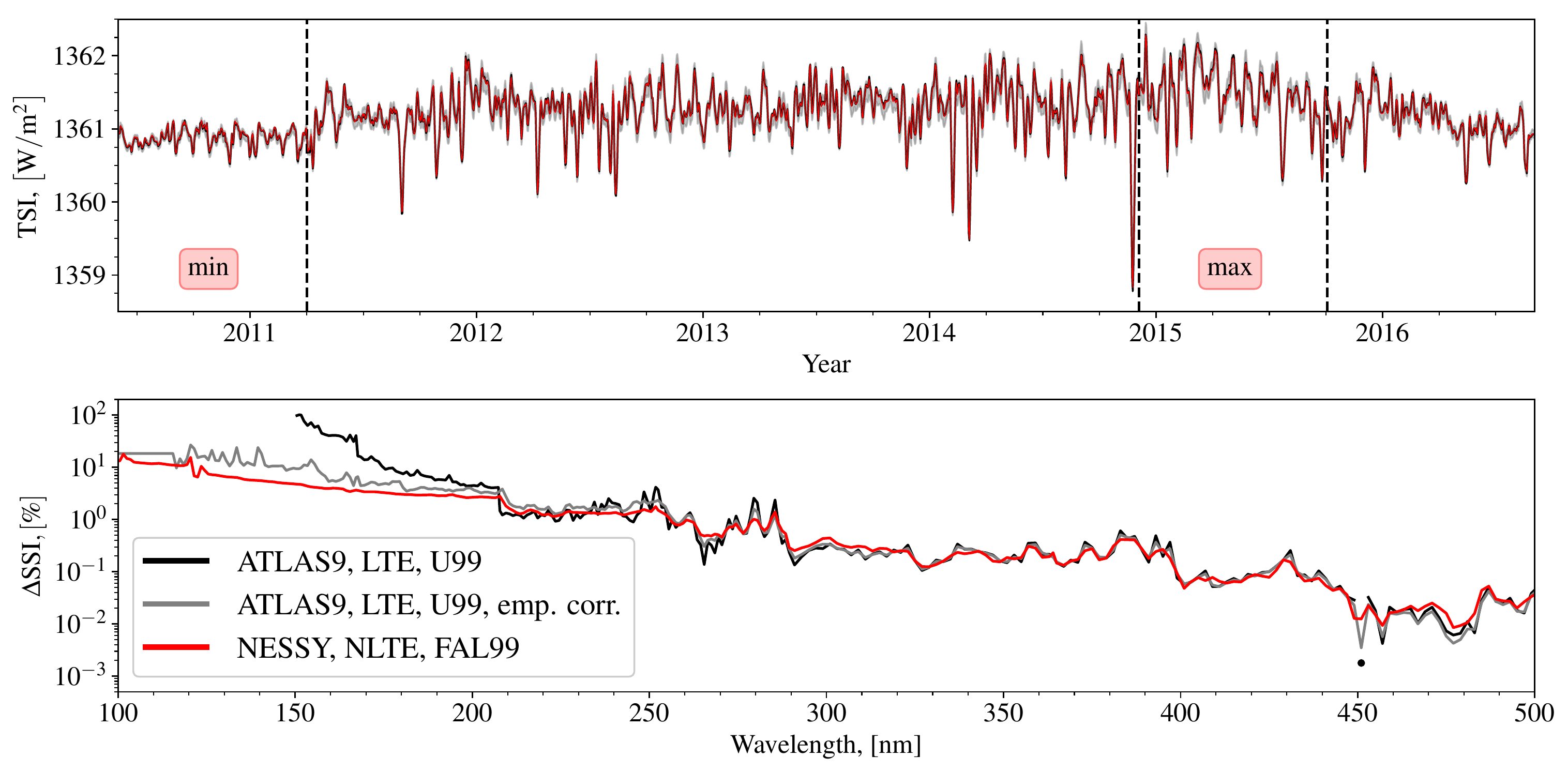}
\caption{ {Top panel:}     {TSI calculated with SATIRE-S using 
                                SDO/HMI filling factors covering the period 
                                of solar cycle 24 in combination 
						    	with spectra computed with NESSY and ATLAS9
						    	(red and black curves, respectively; almost indistinguishable).
						     	TSI 1$\sigma$ uncertainty range
                                as given by the empirically corrected SATIRE-S based on ATLAS9 spectra is shown in gray.
                                The dashed lines designate the periods of minimum 
                                (from 3 June 2010 to 2 April 2011) 
                                and maximum (from 4 December 2014 to 3 October 2015) solar activity 
                                over which the average values of SSI were calculated for
                                computation of SSI variability profile (Eq.~\ref{eq:var_spec}) shown in the bottom panel.}
          {Bottom panel:}  {Relative difference between the SSI at activity maximum and minimum versus wavelength (variability profile, Eq.~\ref{eq:var_spec}).
                                SSI at minimum and maximum solar activity are averages over the periods marked ``min'' and ``max'' (and dashed lines) in the  upper panel.
                                The dot represents a negative value.
                                The resolution of all curves is that of the ATLAS9 spectrum (see Sect.~\ref{desc}).
                                Shortward of about 150 nm the black curve is truncated because of ATLAS9 limitations.
                                }
}
\label{fig:tsi_var_spec}
\end{figure*}

 {Figure~\ref{fig:spec_comp} is concerned with the disc-integrated fluxes.
In Fig.~\ref{fig:clv} we compare ATLAS9 and NESSY quiet Sun 
Centre-to-Limb Variations (CLVs) to the available observations.
We plotted the relative deviations of the ATLAS9 and NESSY CLVs from \cite{neclab1994} fits.
All CLVs have been normalized to the intensity at the centre of the solar disc.
The selection of wavelengths is identical to that used in U99 (see their Fig.~2, bottom panel).
NESSY CLVs are calculated for the cases shown in Fig.~\ref{fig:spec_comp}.
All resulting CLVs show comparable deviations from the fits, 
with NESSY performing somewhat worse at 1047~nm.
Close to the limb the agreement of all CLVs with observations is limited by 
the 1D representation of solar atmosphere.
\citep{koesterke2008, uitcri2011}.}

 {Figure~\ref{fig:contr} shows the comparison of full-disc relative facular, penumbral and umbral contrasts
\begin{equation}\label{eq:rfdc}
C_j(\lambda) = \frac{S_j(\lambda) - S_q(\lambda)}{S_q(\lambda)}, \quad j \in \{f, p, u\},
\end{equation}
calculated with NESSY versus the contrasts calculated with ATLAS9.
Here, $S_q(\lambda)$ is the full disc quiet Sun radiative flux (Eq.~\ref{irr_eq}) and $S_j(\lambda)$ is computed
analogously for each active feature (i.e., as if the quiet Sun or a given active feature occupied the whole solar disc).
Each curve in Fig.~\ref{fig:contr} is derived from the corresponding curves in Fig.~\ref{fig:spec_comp},
so that one can see the impact on the contrasts from the change of the code, radiative transfer treatment and atmospheric model set.}

 {Figure~\ref{fig:contr} shows that the spot contrasts hardly changed. 
The facular contrasts, in turn, differ shortward of the \ion{Al}{I} ionization edge (about 210~nm).
In addition, the NESSY-NLTE-U99 facular contrast longward of 450~nm 
is lower than for the rest of the computations.
Shortward of 210~nm the difference between the facular contrasts 
calculated with the U99 set of models and the FAL99 set 
is due to the absence of chromospheric layers in the U99 models.
Longward of 450~nm the NESSY-NLTE-U99 contrast is shifted downward from its LTE counterpart due to the non-LTE increase of the H$^-$ concentration (see above).
This effect is offset by the presence of chromospheric layers in the FAL99 models. 
Interestingly, all contrasts are remarkably similar in  the 210~nm - 450~nm domain. 
The explanation of this is given in the Appendix.
}

To reconstruct TSI and SSI we substituted the non-LTE NESSY spectra, calculated with the models shown in the upper panel of Fig.~\ref{fig:atm_mod},
and SDO/HMI filling factors covering the period of solar cycle 24 into Eq.~\eqref{irr_eq}. 
Integrating Eq.~\eqref{irr_eq} over wavelength produces the TSI reconstruction shown in the top panel of 
Fig.~\ref{fig:tsi_var_spec} with the red curve. Also shown,  {in black}, are the TSI reconstruction based on the ATLAS9 spectra 
(computed with the U99 set of models) and the empirically corrected SATIRE-S output
\citep[in gray,][see also the Introduction to this paper for a brief description]{yeo2014}.
The TSI computed in these three cases are hardly distinguishable from each other.
We note that the NESSY-NLTE contrasts are slightly higher than the original U99 contrasts in the visible (see Fig.~\ref{fig:contr}), which leads to an increase in their bolometric contrast. As a consequence, the new reconstructions require slightly smaller facular filling factors, so that the fitted value of the $B_\mathrm{sat}$ parameter has increased from 267~G given by \cite{yeo2014} to 346~G.

The lower panel of Fig.~\ref{fig:tsi_var_spec} shows the relative difference between the
average values of SSI over the periods of maximum and minimum activity
\begin{equation}\label{eq:var_spec}
\Delta \mathrm{SSI}(\lambda) = 
\frac{\left<S(t, \lambda)\right>_\mathrm{max} - \left<S(t, \lambda)\right>_\mathrm{min}}
{\left<S(t, \lambda)\right>_\mathrm{min}}.
\end{equation}
Here, $S(t, \lambda)$ is the time-dependent flux as defined by Eq.~\eqref{irr_eq}
and the angular brackets represent the averaging over the minimum and maximum time periods indicated in the top panel of Fig~\ref{fig:tsi_var_spec} (dates are given in the figure caption).

For wavelengths below 210~nm, the non-LTE SSI variability (red) is much closer to the empirically corrected SATIRE-S values (gray) than the LTE results (shown in black). 
Above 210~nm all three approaches yield very similar SSI variability. 
The difference between the non-LTE and LTE SSI variability profiles 
is readily explained in terms of the contrasts shown in Fig.~\ref{fig:contr}. 
Since the transition from ATLAS9 LTE to NESSY non-LTE calculations hardly changes 
the spot contrasts, the difference between the corresponding SSI variabilities is 
determined by the difference of the facular contrasts prevailing below 210~nm.

\section{Summary, conclusions and outlook}\label{conc}
Solar irradiance variability modelling provides important input to models of global climate change \citep{solanki2013, ermolli2013}, 
and increases our understanding of the place of the Sun among other stars in the context of brightness variability patterns \citep[e.g.,][]{basri2013, shapiro2013b, shapiro2014}.
It also provides a foundation for realistic models of stellar brightness variability required in the field of exoplanet detection \citep{boisse2012, korhonen2015}.

The SATIRE-S model has been highly successful in reproducing the observed 
TSI and SSI variability \citep{krivova2003, krivova2006, wenzler2005, wenzler2006, ball2012, yeo2014, yeo2015, yeo2017a}.
However, its output below 300~nm has had to be corrected empirically to overcome the limitations of the LTE approximation in this wavelength range \citep{krivova2006, yeo2014}.
Due to the long-standing debate on the amplitude of the SSI 
variability in the spectral range 200~nm - 400~nm, 
in particular, the disagreement between the empirical and semi-empirical models 
\citep{ermolli2013, thuillier2014b, yeo2015, yeo2017a, egorova2018}, 
and the importance of this range for climate models 
\citep{sukhodolov2014, ball2014b, maycock2015, maycock2016, ball2016}, 
this correction reduces the stringency of SATIRE-S results.

To remove the need for such a correction, we have constructed the non-LTE version of SATIRE-S.
In this version the ATLAS9 \citep{kurucz1992a, caskur1994} LTE spectra of quiet Sun and active components have been replaced with the non-LTE spectra synthesized with the NESSY code \citep{tagirov2017}.
Additionally, for the quiet Sun and facular contrast terms of Eq.~\eqref{irr_eq}, we changed the Kurucz quiet Sun model and the modified FAL93-P model \citep{fontenla1993, unruh1999}, previously used in SATIRE-S, to the corresponding FAL99 models \citep{fontenla1999}.

We analyzed the change of contrasts caused by the use of NESSY and the new set of models (Fig.~\ref{fig:contr})
and compared the non-LTE version of TSI and SSI to the empirically corrected and uncorrected LTE 
ones for solar cycle 24 (Fig.~\ref{fig:tsi_var_spec}).
TSI computed in all three cases are nearly indistinguishable from each other. 
Comparing SSI we found that below 200~nm the non-LTE version agrees well with the empirically  corrected LTE version, 
 {although it does not capture the spectral line structure between 125~nm and 180~nm due to the limitations of the NESSY code
(see the last paragraph of Sect.~\ref{desc})}.
Above 200~nm, where the difference between non-LTE and LTE facular contrasts is not as large as below 200 nm, 
SSI remains essentially unchanged when the non-LTE approach is employed instead of the empirical correction, 
except for the two regions around 300~nm and 470~nm, where the agreement between the two is somewhat worse.
The empirical correction of the SATIRE-S model computed in LTE is thereby shown to give SSI in line with a non-LTE computation.

In this paper we have presented the non-LTE advancement of the SATIRE-S model.
The next step will be the release of non-LTE SATIRE-S with chromospheric lines.
In order for this to be accomplished, the lower variability in Ly$\alpha$ and the absence of chromospheric lines between 125~nm and 160~nm in the NESSY code have to be addressed. 
Apart from that, better 1D semi-empirical models of 
umbra and penumbra with chromospheric layers have to be created.
The recent Atacama Large Millimeter Array (ALMA) observations 
\citep{loukitcheva2017} present an opportunity 
for the development of such models.

\begin{acknowledgements}
We thank Marty Snow for motivating us to carry out this study.
We also thank the International Space Science Institute, Bern, Switzerland for providing financial support and meeting facilities that allowed stimulating discussions (ISSI Team 373 ``Towards a Unified Solar Forcing Input to Climate Studies'').
The research leading to this paper has received funding from STFC consolidated grants ST/N000838/1 and ST/S000372/1, from the European Research Council under the European Union Horizon 2020 research and innovation programme (grant agreement No. 715947) and the German Federal Ministry of Education and Research
(Bundesministerium f{\"u}r Bildung und Forschung) under Project No. 01LG1209A.
The SATIRE-S TSI and SSI reconstructions are available at \url{http://www2.mps.mpg.de/projects/sun-climate/data.html}.
Finally, we thank the referee for their critical assessment of our work 
and the useful suggestions which helped to improve it significantly.
\end{acknowledgements}

\bibliographystyle{aa}
\bibliography{tagirov}

\section*{Appendix: Analysis of the contrast similarity in the 210~nm to 450~nm spectral domain}
\begin{figure}
\centering
\includegraphics[width=9.15cm]{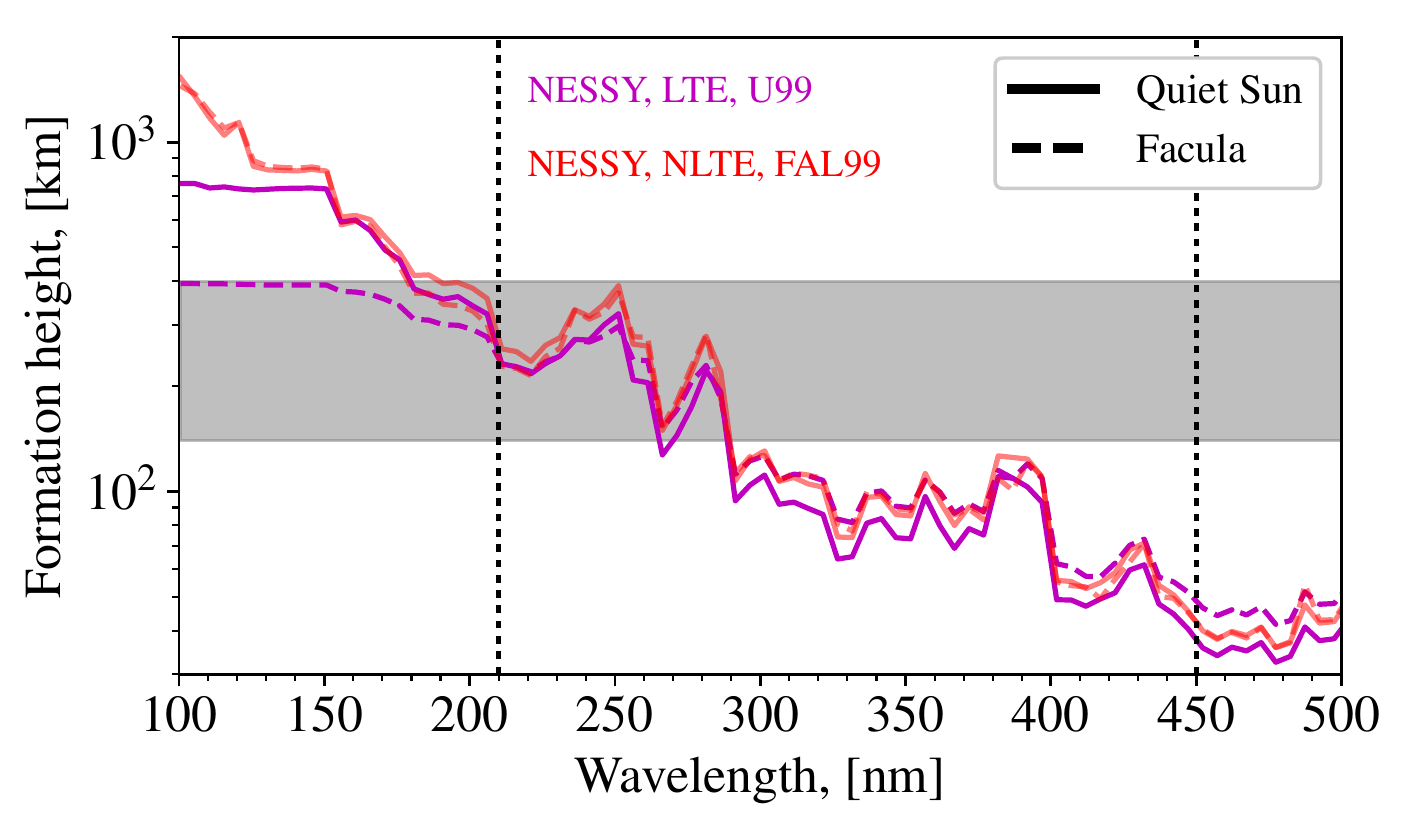}
\caption{ {Formation height of full-disc 
         quiet Sun and facular fluxes computed with NESSY
         for LTE-U99 and NLTE-FAL99 cases.
         The intermediate case (NLTE-U99) has been omitted for visual clarity.
         The vertical dotted lines refer to the same spectral domain as in the top panel of Fig.~\ref{fig:contr}.
         The gray area designates the region, where the temperature differences between quiet Sun and facula in the U99 and FAL99 atmospheric model sets are equal (see Fig.~\ref{fig:atm_mod}, bottom panel). The high resolution formation heights (2000 points per nm) were weighted with the corresponding intensity, averaged over 1~nm intervals and then shifted in the same way as the height grids in Fig.~\ref{fig:atm_mod}.}
		 }
\label{fig:fheight}
\end{figure}

The similarity of the contrasts between 210~nm and 450~nm shown in Fig.~\ref{fig:contr} 
can be interpreted by considering the formation heights of the fluxes. 
Fig.~\ref{fig:fheight} shows 
the formation height of full-disc quiet Sun and facular 
fluxes calculated with NESSY for the LTE-U99 and NLTE-FAL99 cases.
The formation heights were shifted in the same way as the height grids in Fig.~\ref{fig:atm_mod},
so that there is a one-to-one correspondence between the formation heights plotted in Fig.~\ref{fig:fheight} and the temperature structures in Fig.~\ref{fig:atm_mod}.
The idea behind using Figs.~\ref{fig:atm_mod} and \ref{fig:fheight} for the interpretation of the contrasts similarity is to associate the contrasts with the temperature differences at the corresponding formation heights. 
Strictly speaking such an association is only valid under the assumption of LTE, when the  source function is equal to the Planck function and thus is defined by the local  temperature. However, it should be accurate enough for our illustrative purposes. 
In particular, we note that the effect of over-ionization 
on spectral lines only changes their opacity, but not their source function.

 We can see that the formation heights are decreasing and in the NLTE-FAL99 case they are the same for the quiet Sun and faculae almost everywhere.
It means that the NESSY-NLTE-FAL99 contrast is given by the difference between the corresponding quiet-Sun and facular temperatures in Fig.~\ref{fig:atm_mod} at each height.
According to Fig.~\ref{fig:atm_mod}, down to the zero height point, 
this temperature difference and, consequently, the contrast diminish.
In the LTE-U99 case, the formation heights are equal starting from 210~nm up to approximately 300~nm.
In this range they fall into the gray area,
which designates the region of (almost) equal temperature difference between quiet Sun and facula in the U99 and FAL99 sets (see Fig.~\ref{fig:atm_mod}, bottom panel).
Hence, from 210~nm to 300~nm the LTE-U99 formation 
temperature difference and contrast are similar to the NLTE-FAL99 case.
Between 300~nm and 450~nm the spectrum forms in the regions where the U99 temperature difference between quiet Sun and facula decreases faster than in the FAL99 case.
However, the formation height in the U99 facula becomes greater than in the quiet Sun, thus compensating for the faster decrease and preserving the equality of LTE-U99 and NLTE-FAL99 contrasts up to 450~nm.

\end{document}